\documentclass[conference,10pt]{IEEEtran}
\IEEEoverridecommandlockouts
\usepackage{cite}
\usepackage{hyperref}
\usepackage{url}
\usepackage{pifont}
\usepackage{graphicx}
\usepackage{subfigure}
\usepackage{float}
\usepackage{stfloats}  
\usepackage{threeparttable}
\usepackage{booktabs}
\usepackage{multirow}
\usepackage{makecell}
\usepackage{caption}  
\usepackage[cmex10]{amsmath} 
\usepackage{amssymb}  
\usepackage{bm}       
\usepackage{mathrsfs} 
\usepackage{amsthm}   

\usepackage[linesnumbered,ruled,vlined]{algorithm2e}
\usepackage{xcolor}      

\usepackage{array}
\usepackage{mdwmath}     
\usepackage{mdwtab}      
\usepackage{eqparbox}    
\usepackage{etoolbox}    
\usepackage{supertabular} 

\usepackage{flushend}

\usepackage[acronym]{glossaries}
\newacronym{6g}{6G}{sixth-generation}
\newacronym{5g}{5G}{fifth-generation}
\newacronym{iot}{IoT}{Internet-of-Things}
\newacronym{aiot}{AIoT}{artificial intelligence-of-things}
\newacronym{ai}{AI}{artificial intelligence}
\newacronym{mllm}{MLLM}{multimodal large language model}

\newcommand{\e}{\begin{equation}}
\newcommand{\ee}{\end{equation}}
\newcommand{\eqn}{\begin{eqnarray}}
\newcommand{\eeqn}{\end{eqnarray}}

\makeatletter
\renewcommand\normalsize{\@setfontsize\normalsize{10pt}{12pt}}
\makeatother

\begin{document}

\title{
Token-Domain Multiple Access: Exploiting \\
\vspace{-1mm}
Semantic Orthogonality for Collision Mitigation
\vspace{-4mm}
}
\author{
\IEEEauthorblockN{Li Qiao\IEEEauthorrefmark{1}\IEEEauthorrefmark{3}, Mahdi Boloursaz Mashhadi\IEEEauthorrefmark{3}, Zhen Gao\IEEEauthorrefmark{2}\IEEEauthorrefmark{1}, and Deniz Gündüz\IEEEauthorrefmark{4}}
\IEEEauthorblockA{\IEEEauthorrefmark{1}School of Information and Electronics, Beijing Institute of Technology, Beijing 100081, China}
\IEEEauthorblockA{\IEEEauthorrefmark{2}Advanced Technology Research Institute, Beijing Institute of Technology (Jinan), Jinan 250307, China}
\IEEEauthorblockA{\IEEEauthorrefmark{3}5GIC \& 6GIC, Institute for Communication Systems (ICS), University of Surrey, GU2 7XH Guildford, U. K.}
\IEEEauthorblockA{\IEEEauthorrefmark{4}Department of Electrical and Electronic Engineering, Imperial College London, London SW7 2AZ, U. K.}
Email: \{qiaoli, gaozhen16\}@bit.edu.cn, {m.boloursazmashhadi}@surrey.ac.uk, d.gunduz@imperial.ac.uk
\vspace{-6mm}
\thanks{Z. Gao was supported in part by the NSFC under Grant U2233216 and Grant 62471036, in part by Beijing Natural Science Foundation under Grant L242011, in part by Shandong Province Natural Science Foundation under Grant ZR2022YQ62, and in part by Beijing Nova Program. D. Gunduz received funding from UKRI for the project AI-R (ERC Consolidator Grant, EP/X030806/1) and the SNS JU project 6G-GOALS under the EU’s Horizon program (Grant Agreement No. 101139232).}
}


\maketitle

\vspace{-7mm}
\begin{abstract}
\textit{Token communications} is an emerging generative semantic communication concept that reduces transmission rates by using context and transformer-based \textit{token} processing, with tokens serving as universal semantic units. In this paper, we propose a semantic multiple access scheme in the token domain, referred to as \textit{ToDMA}, where a large number of devices share a tokenizer and a modulation codebook for source and channel coding, respectively. Specifically, the source signal is tokenized into sequences, with each token modulated into a codeword. Codewords from multiple devices are transmitted simultaneously, resulting in overlap at the receiver. The receiver detects the transmitted tokens, assigns them to their respective sources, and mitigates token collisions by leveraging context and semantic orthogonality across the devices' messages. Simulations demonstrate that the proposed ToDMA framework outperforms context-unaware orthogonal and non-orthogonal communication methods in image transmission tasks, achieving lower latency and better image quality.
\end{abstract}


\IEEEpeerreviewmaketitle
\vspace{-4mm}
\section{Introduction}

The emergence of \gls*{mllm}, such as GPT-4 Omni \cite{openai_2024_gpt4o} and Show-o \cite{xie2024show}, represents a major advancement in artificial intelligence (AI), combining large language models (LLMs) with the capability to handle diverse data modalities like text, images, video, and audio \cite{caffagni2024r}. At the heart of \gls*{mllm}s is the transformer architecture, which processes {\it tokens} using self-attention \cite{vaswani2017attention}. The transformation of signals into tokens, known as {\it tokenization}, uses a {\it tokenizer}, typically a learned codebook, to map input segments into discrete tokens. This process breaks complex data into manageable parts, as seen in WordPiece tokenizers for text \cite{devlin2018bert} and VQ-VAE tokenizers for images \cite{van2017neural}. 

Generative transformer models rely on self-supervised pre-training over vast datasets of tokens. For example, models like BERT \cite{devlin2018bert} use {\it masked language modeling}, where random tokens are masked during training, and the model predicts the masked tokens using bidirectional context. Similarly, masked image modeling, e.g., MaskGIT \cite{chang2022maskgit}, has shown success in vision tasks. {\it Next-token prediction} approach is another effective pre-training task, as demonstrated by GPT models \cite{openai_2024_gpt4o} for text generation and VQ-GAN \cite{esser2021taming} for high-quality image synthesis.

In this paper, we propose \textit{token communications} (TokCom), a framework that uses tokens as semantic content for future wireless networks. For example, a corrupted TokCom packet might result in an incomplete message like, ``The cat is [MASK] on the mat.", where [MASK] represents a lost token. A pre-trained masked language model such as BERT can predict the missing word/token, e.g., ``sitting" in the above example, using context to avoid retransmissions. Building upon this idea, we explore the following key questions: What does the multi-access architecture of TokCom look like, and how can contextual information be effectively utilized to enhance communications? To address these, we introduce the concept of token-domain {\it semantic orthogonality} as a novel dimension for multiple access. This leads to the proposal of the {\it token-domain multiple access (ToDMA)} scheme, where multiple devices can transmit over the same wireless multiple access channel (MAC) in a non-orthogonal manner, leading to overlapping signals. An \gls*{mllm} at the receiver separates the devices' individual signals at the token level, leveraging semantic orthogonality and contextual information. 

For instance, the sentences ``The cat jumped over the high fence.” and ``Books provide endless inspiration and knowledge.” are semantically orthogonal, enabling a model like GPT-4 to reconstruct them from mixed tokens/words. However, 
as the number of devices and the dimension of their signals increase, the prediction space for \gls*{mllm}s grows exponentially, making the complexity of relying solely on \gls*{mllm}s impractical. To tackle these challenges, this paper proposes a practical non-orthogonal multiple access scheme that leverages the advantages provided by transformer-based token processing, utilizing {semantic orthogonality} to enhance next-generation communication technologies.

\vspace{-2mm}
\section{Related Works}

\subsection{Generative AI Meets Semantic Communications}
Goal-oriented and semantic communications have demonstrated significant potential to transform next-generation wireless networks using AI, making them more efficient, timely, and intelligent \cite{strinati2024goal, beyondBits}. 
With advancements in generative AI (GenAI), semantic communications can be enhanced by utilizing GenAI models to improve the comprehension and generation of high-fidelity communication content, a concept recently proposed as {\it generative semantic communications} \cite{Liang2024generative,DaiMagzine}.

GenAI has demonstrated significant potential to enhance point-to-point semantic communications across various dimensions, including source coding \cite{deletang2023language, qiao2024latency, cicchetti2024language, ren2024generative,Extreme}, channel coding \cite{choukroun2022error,nam2024language}, and deep joint source-channel coding (DeepJSCC) \cite{yilmaz2024high, erdemir2023generative,devoto2024adaptive,zhang2025semantics}. Recent studies \cite{yilmaz2023distributed, liang2024orthogonal, zhang2023deepma, mu2023exploiting} have integrated DeepJSCC with non-orthogonal multiple access (NOMA), achieving impressive bandwidth savings especially in two-user scenarios. However, scaling these solutions to a massive number of devices presents challenges due to the complexity of end-to-end neural network training. Furthermore, the integration of powerful GenAI models, e.g., the potential of \gls*{mllm}s for token prediction based on multimodal context, into semantic multiple access schemes remains an under-explored area.

\subsection{Evolution of Multiple Access in 6G}

Advancements in multiple access towards 6G focus on efficient random access for numerous uncoordinated devices with sporadic activity and short data packets \cite{ding2024next,10130590,10381510}. Key 6G technologies like NOMA and grant-free random access (GFRA) are expected to meet these demands \cite{liva2024unsourced}. For example, in GFRA, after receiving a beacon signal from the base station (BS), devices transmit their unique preambles over the same time-frequency resources. The BS then performs device activity detection, channel state information (CSI) estimation, and data detection. This protocol is commonly utilized in studies such as \cite{Liuliang, qiao2023sensing, wang2024covariance}.
To accommodate a larger number of devices, unsourced massive access (UMA) introduces a shared preamble codebook, where messages are represented as indices of codewords within the codebook. The BS focuses on decoding the transmitted codewords from the shared codebook, without necessarily associating them to their transmitters \cite{polyanskiy2017perspective, shyianov2020massive, tian2024design}. 

As devices become increasingly intelligent and computationally capable, 6G is expected to evolve toward the artificial intelligence of things (AIoT) paradigm, characterized by the deep integration of AI with massive communication protocols. For instance, the UMA protocol was specifically designed for efficient computation in \cite{qiao2024massive}, enabling communication-efficient distributed learning and inference. On the other hand, generative semantic communications can play a pivotal role in significantly reducing the latency in massive communication systems, serving as a key motivation for our work.

\begin{figure*}[t]
     \centering
     \includegraphics[width = 1.8\columnwidth,keepaspectratio]{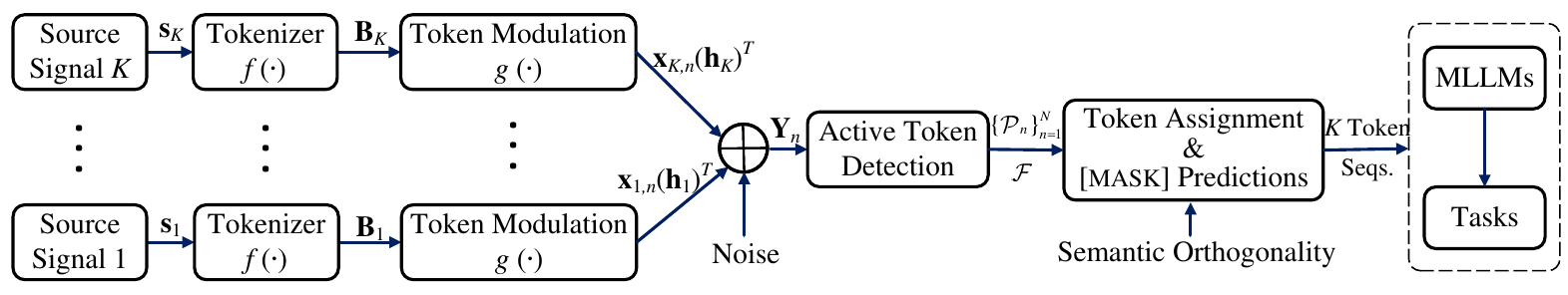}
     \captionsetup{font={footnotesize, color = {black}}, singlelinecheck = off, justification = raggedright,name={Fig.},labelsep=period}
     \caption{The proposed Token-Domain Multiple Access (ToDMA) framework.}
     \label{fig2}
     \vspace{-7mm}
\end{figure*}

\section{Contributions}
To tackle these challenges, we summarize our main contributions as follows:
\begin{itemize} 

\item {\bf Proposed ToDMA framework}: To the best of our knowledge, this is the first paper to propose a multiple access scheme in the token domain. Specifically, the proposed ToDMA framework tokenizes source signals using pre-trained models, modulates each token to a codeword from an orthonormal modulation codebook and allows overlapped codeword transmissions from multiple devices. The non-orthogonal and grant-free uplink transmission is able to significantly reduce signaling overhead and latency of massive communications.
\item {\bf Using semantic orthogonality for collision mitigation}: The receiver initially detects the transmitted tokens along with their corresponding CSI, and assigns each token to a source with matching CSI. However, non-orthogonal transmission may lead to token collisions, resulting in some detected tokens remaining unassigned and certain positions in the reconstructed token sequences left unfilled. To address this, we propose leveraging pre-trained \gls*{mllm}s to associate unassigned tokens with the masked positions of individual sources by utilizing contextual information and semantic orthogonality. Simulation results confirm that ToDMA achieves reduced communication latency for image transmission scenarios involving a large number of potential devices.

\end{itemize}

\textit{Notation}: Boldface lower and upper-case symbols denote column vectors and matrices, respectively. For a matrix ${\bf A}$, ${\bf A}^T$, ${\bf A}^H$, $[{\bf{A}}]_{m,n}$ denote the transpose, Hermitian transpose, and the $m$-th row and $n$-th column element of ${\bf{A}}$, respectively.  $[{\bf{A}}]_{n,:}$ denotes the $n$-th row of ${\bf A}$. $|\mathcal{P}|$ denotes the cardinality of the set $\mathcal{P}$, and ${\bf 0}_{m}$ is a vector of all zeros with dimensions ${m}$. $\mathbf{I}_n$ denotes identity matrix with size $n$.

\section{Proposed Token-Domain Multiple Access (ToDMA) Framework}

In this section, we first introduce the transmitter architecture for our proposed ToDMA framework. Then, we provide the corresponding multiple access signal model.

\subsection{Transmitter Design}
\label{Sec:Transmitter}
Fig. \ref{fig2} depicts the general block diagram of the proposed ToDMA framework. We consider a general multiple access scenario, where $K_{\rm T}$ single-antenna devices are served by a BS equipped with $M$ antennas. Despite a large total number of devices, typically only a small fraction, $K$ (where $K \ll K_T$), have data to transmit at a given time \cite{liva2024unsourced}. We denote the source signal of the $k$-th device as a vector ${\bf s}_k$, $\forall k\in[K]$, which can represent the data of any modalities, like text, vision, audio, etc. In the following, we will describe the tokenization and token modulation processes at each device.

\subsubsection{Tokenization}
Mathematically, a tokenizer cuts up the original signal into smaller pieces, each belonging to a prescribed finite alphabet. Eventually the signal is mapped to a sequence of indices from this alphabet. We use $f(\cdot)$ to represent the tokenizer and $Q$ to represent the size of the {\it token codebook}, therefore the {\it tokenization} process can be denoted as ${\bf B}_k = f({\bf s}_k) \in \{0, 1\}^{Q\times N}$, where the $n$-th column of ${\bf B}_k$, denoted by ${\bf b}_{k,n}$, is a one-hot vector representing the corresponding token. Without loss of generality, the length of resulting token sequences $N$ is considered the same for all the devices. Later, we will use $f^{-1}(\cdot)$ to denote the de-tokenization process.

\subsubsection{Token Modulation}

Assuming that the same tokenizer is shared among the devices, we propose to use a shared modulation codebook for multiple access, denoted as ${\bf U} \in \mathbb{C}^{L\times Q}$, 
where $L$ denotes the length of each codeword. For simplicity, we consider the columns of ${\bf U}$ are orthonormal vectors, i.e., ${\bf U}^H{\bf U}=\mathbf{I}_Q$. We use $g(\cdot)$ to represent the token modulation process; hence, the output is denoted by
${\bf X}_k = {\bf U}{\bf B}_k = g({\bf B}_k) \in \mathbb{C}^{L\times N}$, where each token is modulated to a modulation codeword. We denote the $n$-th column of ${\bf X}_k$ as ${\bf x}_{k,n}= {\bf U}{\bf b}_{k,n} \in\mathbb{C}^{L}$. The advantages of the proposed token modulation are mainly two fold: \ding{172} Shared modulation codebook enables uncoordinated random access of massive number of devices, which reduces the access latency and is easy for implementation. \ding{173} The basic communication unit changes from the conventional ``bits" to ``tokens", which facilitates applying transformer-based token processing and \gls*{mllm}s at the receiver.

\subsection{Multiple Access Signal Model}

In a typical massive communication scenario envisioned for future 6G networks, BSs serve a large number of uncoordinated devices \cite{liva2024unsourced}. Periodically, the BS broadcasts beacon signals. If a device has data to send on the uplink, it transitions from idle to active state without the BS knowing which devices are active at any given moment. Upon receiving the beacon signal, these active devices immediately transmit via the MAC. We consider that $K$ active devices use the same $N$ communication resources, e.g., $N$ time slots, each consisting of $L$ symbols, to transmit their individual $N$ tokens. In the $n$-th time slot, $\forall n\in[N]$, the transmitted signals are overlapped at the receiver, i.e., the received signal ${\bf Y}_{n}\in\mathbb{C}^{L\times M}$ at the receiver is expressed as
\vspace{-2mm}
\begin{align}\label{MA-SignalMod}
    {\bf Y}_{n}&=\sum\limits_{k\in[K]} {\bf x}_{k,n} \left({\bf h}_k\right)^T + {\bf Z}_{n} \nonumber \\
    &={\bf U} \sum\limits_{k\in[K]}{\bf b}_{k,n} \left({\bf h}_k\right)^T + {\bf Z}_{n} = {\bf U}{\bf H}_n + {\bf Z}_{n},
    \vspace{-2mm}
\end{align}
where ${\bf h}_k\in \mathbb{C}^{M}$ denotes the channel vector between the $k$-th device and the BS, ${\bf H}_n = \sum\nolimits_{k\in[K]}{\bf b}_{k,n} ({\bf h}_k)^T \in\mathbb{C}^{Q\times M}$ is termed as an {\it equivalent channel matrix}, and the noise matrix ${\bf Z}_{n}$ follows an independent and identically distributed (i.i.d.) complex Gaussian distribution with zero mean and variance $\sigma^2{\bf I}_n$. 
Without loss of generality, the channel vector ${\bf h}_k$, $\forall k$, is assumed to remain the same throughout the $N$ time slots.

\vspace{-2mm}
\section{Receiver Design: Token Detection, Token Assignment, and Masked Token Predictions}

\subsection{Problem Formulation}
The receiver's objective is to reconstruct the $K$ transmitted token sequences from ${\bf Y}_{1},...,{\bf Y}_{N}$, i.e., to estimate the token sequences $\{{\bf B}_k, k \in [K]\}$. This process can be divided into several steps. First, the receiver identifies which tokens are transmitted at each time slot. Based on (\ref{MA-SignalMod}), the task of {\it token detection} can simultaneously determine the CSI associated with each transmitted token. The next step involves assigning detected tokens to their respective devices using the CSI. If \textit{token collisions} occur, where several devices transmit the same token, certain tokens cannot be assigned to any devices, resulting in masked tokens at specific positions. Finally, we propose utilizing a bi-directional transformer to predict these masked tokens by leveraging {\it semantic orthogonality} and {\it contextual information}.

\subsection{Active Token Detection}

Given that the columns of modulation codebook ${\bf U}$ is orthonormal vectors, the receiver obtains an estimate of ${\bf H}_n$ by applying projection using the orthonormal basis vectors as follows
\vspace{-2mm}
\begin{align}\label{MA-Receiver}
    \widehat{\bf H}_n = {\bf U}^H{\bf Y}_{n}&= {\bf H}_n + {\bf U}^H{\bf Z}_{n},
\end{align}
where ${\bf U}^H{\bf Z}_{n}$ does not change the noise variance since ${\bf U}^H{\bf U}=\mathbf{I}_Q$. 
Given any time slot \( n \in [N] \), we denote \( \mathcal{P}_n \) and \( \mathcal{F}_n \) as the indices and the values of the rows in \( \widehat{\bf H}_n \) whose energy normalized by \( M \) exceeds a threshold \( T_h \), respectively.
In other words, $\mathcal{P}_n$ and $\mathcal{F}_n$ represent the estimated tokens and the related channel vectors at the $n$-th time slot, respectively. Specifically, we have $\mathcal{F}_n= \{{\bf h}_{\phi, n} \big| {\bf h}_{\phi, n} = [\widehat{\bf H}_n]_{\phi, :}, \phi\in \mathcal{P}_n\}$.

\begin{figure}[t]
    \centering
    \captionsetup{font={footnotesize, color={black}}, singlelinecheck=off, justification=raggedright, name={Fig.}, labelsep=period}

    \subfigure[Token assignment of a given time slot]{
        \includegraphics[width=0.77\columnwidth, keepaspectratio]{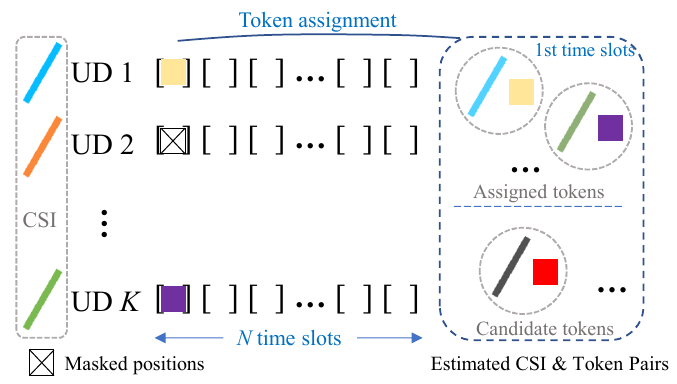}
        \label{figTokMatch1}
    }

    \vspace{-3mm} 

    \subfigure[Masked tokens prediction of a given device]{
        \includegraphics[width=0.77\columnwidth, keepaspectratio]{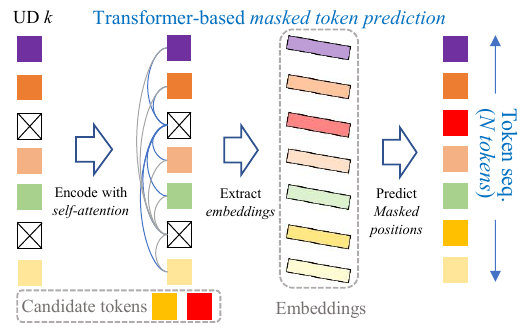} 
        \label{figTokMatch2}
    }
\vspace{-3mm}
    \caption{Illustration of {\it token assignment} and {\it masked token prediction}.}
    \label{figTokMatch}
    \vspace{-6mm}
\end{figure}

\subsection{Token Assignment}
\label{Sec:tokenAssign}
Here, the CSI acts as a matching anchor that links detected tokens to their respective devices. We assume the CSI of $K$-devices are known, denoted by $\{{\bf h}_1, {\bf h}_2,...,{\bf h}_K\}$. We initialize the estimated token sequences as ${\bf c}_k={\bf 0}_{N}$, $\forall k\in[K]$. Note that the elements of ${\bf c}_k$ shall be integers within $[Q]$ to represent tokens, which is equivalent to the one-hot vector representation introduced in Section \ref{Sec:Transmitter}. The token assignment process is described as follows. \ding{172} {\bf Initial Assignment}: We define a {\it residual token set} $\mathcal{P}_n^r$, which is initialized as $\mathcal{P}_n^r=\mathcal{P}_n$, $\forall n\in[N]$. We iterate $k$ from $1$ to $K$, in each iteration, we first find the token that minimize the Euclidean distance:
    $$\phi^{\star}_n={\rm arg\mathop{min}\nolimits}_{\phi\in \mathcal{P}_n}\|{\bf h}_k-{\bf h}_{\phi,n}\|_2, n\in [N],$$
where $\|\cdot\|_2$ denotes the $l_2$ norm. Next, if and only if $\|{\bf h}_k^0-{\bf h}_{\phi^{\star}_n,n}\|_2^2/M<T_h$, we assign token $\phi^{\star}_n$ to $[{\bf c}_k]_n$, and then update $\mathcal{P}_n^r$ by deleting $\phi^*_n$; Otherwise, we assign ``[MASK]" to $[{\bf c}_k]_n$. \ding{173} {\bf Fine-Grained Update}: After the {\it initial assignment}, if $\mathcal{P}_n^r=\emptyset$, there is no token collision in this time slot; if $|\mathcal{P}_n^r|=1$, we assign the only residual token in $\mathcal{P}_n^r$ to the ``[MASK]" positions of $[{\bf c}_k]_n$, $\forall k\in[K]$; If $|\mathcal{P}_n^r|>1$, the elements of this residual token set serve as {\it candidate tokens} to assist the following steps.

Fig. \ref{figTokMatch1} illustrates the token assignment process. It can be seen that by leveraging CSI, the receiver can accurately associate transmitted tokens with their originating devices. If token collisions occur, we get [MASK] in the estimated token sequence and corresponding {\it candidate tokens}. Note that the selection of threshold $T_h$ is based on the noise variance.

\subsection{Masked Token Predictions}
\label{Sec:tokenPredict}
Thanks to the advancements in GenAI, especially the invention of the transformer architecture, the complex relationships among tokens in the whole context can be learned and inferred efficiently at test time. Hence, iterating from $1$ to $K$, in each iteration, we use a pre-trained bi-directional transformer to predict the masked tokens exploiting the contextual information and the {\it candidate tokens}. For example, if ``$[{\bf c}_k]_n$=[MASK]" and $|\mathcal{P}_n^r|>1$, the pre-trained transformer predicts the best choice from among the {\it residual token set} $\mathcal{P}_n^r$ exploiting the contextual tokens. Note that by exploiting the {\it residual token set}, the search space can be significantly reduced from $Q$ to $|\mathcal{P}_n^r|$, which improves the prediction accuracy. 

Fig. \ref{figTokMatch2} presents the process of masked token prediction. The input token sequence, including the masked tokens, is encoded using the pre-trained transformer architecture, which leverages the bi-directional self-attention mechanism to capture contextual relationships among the tokens. The transformer outputs embeddings for each token in the sequence. Based on the extracted embeddings and context, the model predicts the values of the masked tokens, filling in the gaps to complete the token sequence.

\begin{algorithm}[t]
\footnotesize 
\caption{Proposed ToDMA Framework}
\label{Algo:Receiver}
\SetAlgoLined
\KwIn{Source signal ${\bf s}_k$, $k\in[K]$, modulation codebook ${\bf U}$, tokenizer/de-tokenizer $f(\cdot)/f^{-1}(\cdot)$, noise variance $\sigma^2$.}
\KwOut{Token sequences \{${\bf c}_k\in\{ [Q]\}^N$, $\forall k\in[K]$\}.}

\tcp{Transmitters}

\textbf{Tokenization:} ${\bf B}_k = f({\bf s}_k) \in \{0, 1\}^{Q\times N}, \forall k$\;
\label{Alg1:tokenization}
\textbf{Token Modulation:} ${\bf X}_k = {\bf U}{\bf B}_k = g({\bf B}_k) \in \mathbb{C}^{L\times N}$\;
\label{Alg1:modulation}
\textbf{Multiple Access:} After receiving a beacon signal, active devices transmit their modulation codewords simultaneously, as described in (\ref{MA-SignalMod})\; 
\label{Alg1:MAC}
\tcp{Receiver}
\textbf{Token Detection:} Acquiring $\mathcal{P}_n$ and $\mathcal{F}_n$, $\forall n$, according to (\ref{MA-Receiver}) \;
\label{Alg1:TokDetect}
\textbf{Token Assignment:} Using the estimated CSI $\mathcal{F}_n$ to assign tokens $\mathcal{P}_n$ to the respective devices, according to Section \ref{Sec:tokenAssign}\;
\label{Alg1:Assign}
\textbf{Masked Token Prediction:} Using pre-trained bi-directional Transformer to predict masked positions, according to Section \ref{Sec:tokenPredict}\;
\label{Alg1:MTP}
\textbf{Return:} \{${\bf c}_k$, $\forall k\in[K]$\}\;
\label{Alg1:Output}
\end{algorithm}

The proposed ToDMA framework is outlined in {\bf Algorithm \ref{Algo:Receiver}}. The output consists of the estimated tokens, which can either be passed to de-tokenizers for reconstruction or used as input for MLLMs to perform downstream tasks. In our simulation, we focus solely on reconstruction, leaving other applications for future exploration.

\begin{figure*}[t]
\centering
\subfigure[]{
    \begin{minipage}[t]{0.33\linewidth}
        \centering
\label{fig:TPe}
        \includegraphics[width = 0.95\columnwidth,keepaspectratio]{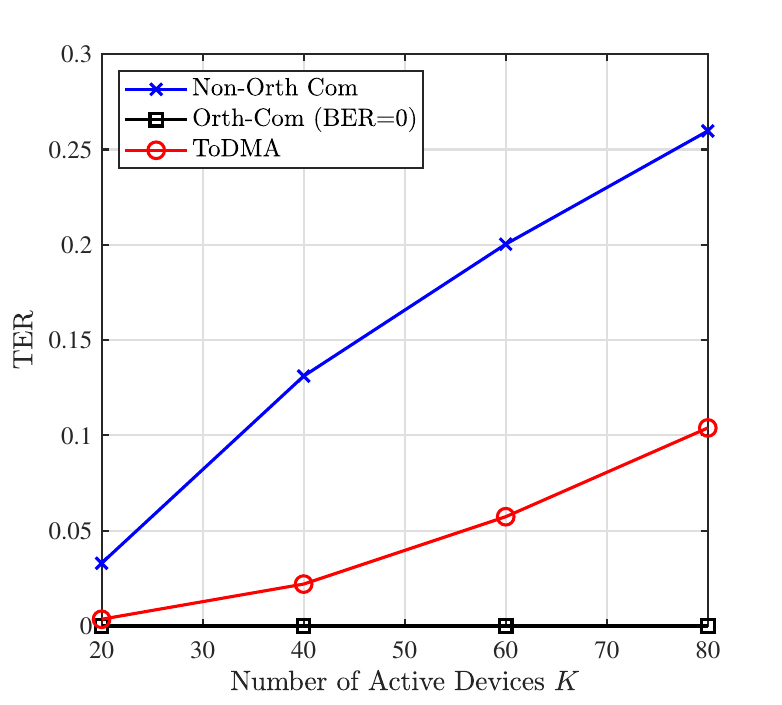}\\
    \end{minipage}%
}%
\subfigure[]{
    \begin{minipage}[t]{0.33\linewidth}
        \centering
\label{fig:TSER}
        \includegraphics[width = 0.95 \columnwidth,keepaspectratio]{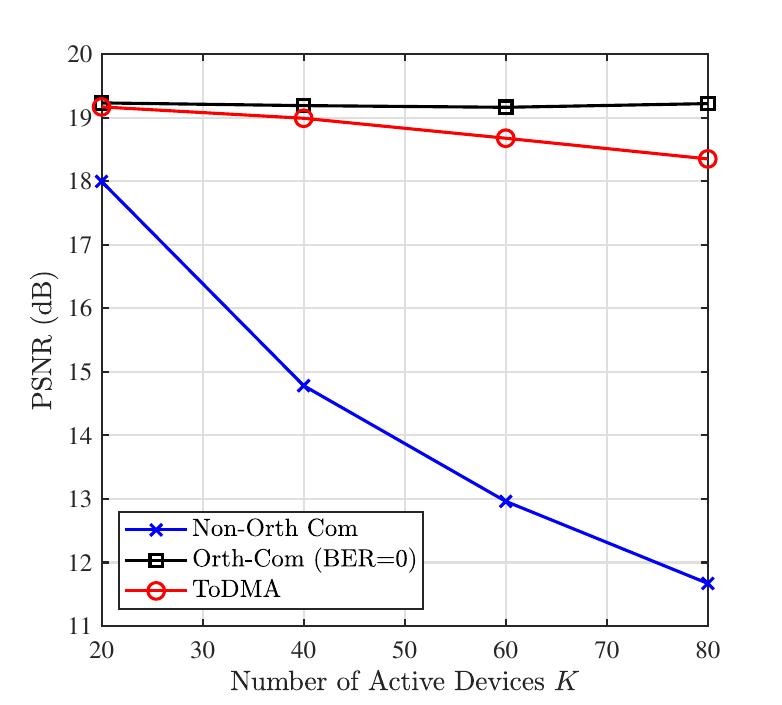}\\
    \end{minipage}%
}%
\subfigure[]{
    \begin{minipage}[t]{0.33\linewidth}
        \centering
\label{fig:TBER}
        \includegraphics[width = 0.95 \columnwidth,keepaspectratio]{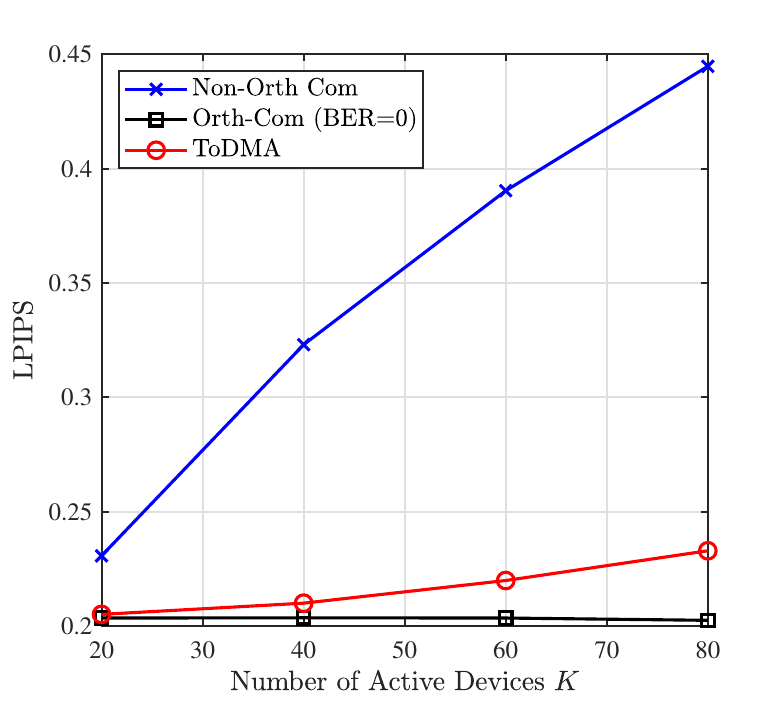}\\
    \end{minipage}%
}%
\centering
\setlength{\abovecaptionskip}{-1mm}
\captionsetup{font={footnotesize}, singlelinecheck = off, justification = justified,name={Fig.},labelsep=period}
\caption{Performance comparison of the proposed ToDMA-based image transmission with benchmark schemes, validated on the ImageNet100 dataset, SNR$=25$\,dB. The benchmark ``Orth-Com" is evaluated with ${\text{BER}=0}$. (a) TER performance versus the number of active devices $K$; (b) PSNR performance versus $K$ (the higher the better); (c) LPIPS performance versus $K$ (the lower the better).}
\label{Sim:Perf_Ka}
\vspace{-7.5mm}
\end{figure*}

\vspace{-3mm}
\section{Simulation Results}
\label{Sec:Simul}

In this section, we have conducted extensive simulations on image transmission to evaluate the performance of the proposed ToDMA framework.
\subsection{Parameter settings}
{\bf Dataset:} We consider the well-known benchmarking dataset ImageNet-100 \cite{russakovsky2015imagenet}. Specifically, we consider 100 classes and each class has 1300 images with resolutions $256\times 256$. In each Monte-Carlo simulation, active device randomly chooses one image from the dataset.
{\bf Tokenizer:} We employ a pre-trained VQ-GAN-based tokenizer provided by \cite{esser2021taming}. The tokenizer features a codebook size of $Q=
1024$, with each image encoded as a sequence of $N=256$ tokens. 
{\bf Masked image predictions:} We employed the pre-trained vision transformer (ViT) model described in \cite{chang2022maskgit}.

\subsubsection{Wireless Parameters}
We consider a massive access scenario with $K_{\rm T}=500$ devices, while the number of active devices $K$ is much smaller than $K_{\rm T}$. 
The number of receive antennas at the BS is $M=256$. The channel vectors ${\bf h}_k$, $\forall k\in[K]$, are generated according to Rayleigh fading. The threshold $T_h$ is set to $2\sigma^2$. We use a normalized discrete Fourier transform codebook ${\bf U} \in \mathbb{C}^{L \times Q}$, so that the orthogonality between columns is ensured, with $L = Q$. The signal-to-noise ratio (SNR) is set to 25\,dB.

\subsubsection{Comparison schemes}
We consider an orthogonal frequency division multiplexing (OFDM) system. The following benchmarks are considered: 
{\bf Context-unaware non-orthogonal scheme}: 
    This benchmark is modified from the proposed ToDMA framework, i.e. the transmitter is the same as ToDMA, while the receiver only runs lines \ref{Alg1:TokDetect} and \ref{Alg1:Assign} of the proposed Algorithm \ref{Algo:Receiver}, without using the contextual information. If there exist [MASK] positions after the assignment, it randomly selects a token from the token codebook for each [MASK] position. We call this scheme as ``{\it Non-Orth Com}". {\bf Context-unaware orthogonal scheme}: Tokens are converted to bits and transmitted using an uncoded adaptive quadrature amplitude modulation (QAM) scheme \cite{goldsmith1997variable} with desired bit error rate (BER). The entire bandwidth is equally assigned to the $K_{\rm T}$ devices in this benchmark; hence, the transmissions of multiple devices are orthogonal. We denote this scheme as ``{\it Orth-Com}".

\subsubsection{Token Communication Metrics}
We define token error rate (TER) as
$\text{TER} = \frac{1}{2NK}\sum_{k=1}^{K}\|\widehat{\bf B}_k - {\bf B}_k\|_0.$ It is worth emphasizing that TER is a general metric for ToDMA without specific data modalities. To evaluate the image transmission performance, we consider the well-known PSNR and LPIPS to measure the distortion and perception qualities, respectively.

\begin{figure}[t]
     \centering
     \includegraphics[width = 0.85 \columnwidth,keepaspectratio]{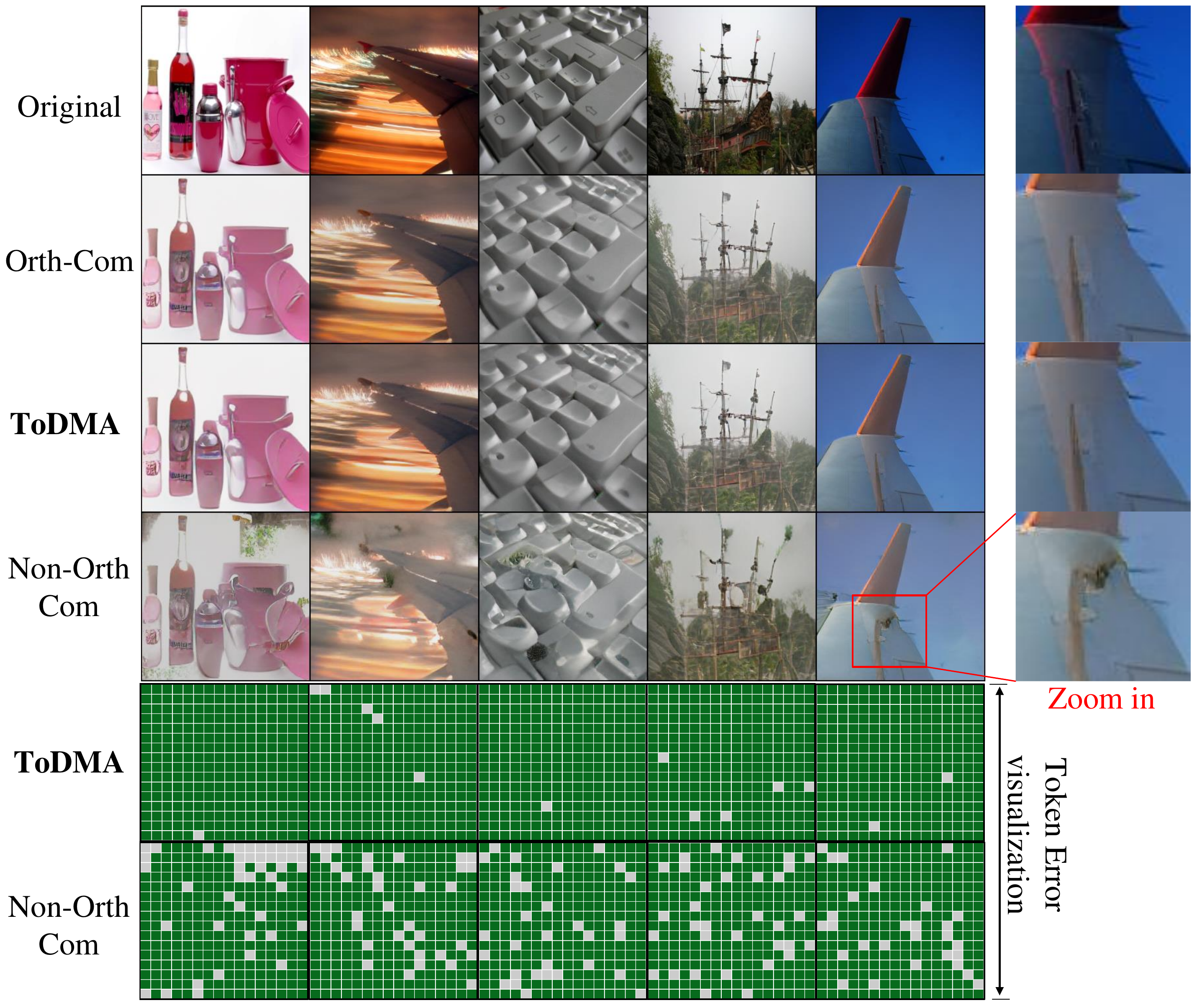}
     \captionsetup{font={footnotesize, color = {black}}, singlelinecheck = off, justification = justified,name={Fig.},labelsep=period}
     \caption{Visual quality illustration for $K=40$. Due to space limitations, only 5 images out of the 40 devices/images are shown, without cherry-picking. From top to bottom: Original images; reconstructed images using ``Orth-Com"; reconstructed images using ``ToDMA"; reconstructed images using ``Non-Orth Com"; token error visualization for ``ToDMA"; and token error visualization for ``Non-Orth Com". 
     }
     \label{Sim:Visual-1}
     \vspace{-7mm}
\end{figure}

Fig. \ref{Sim:Perf_Ka} shows the TER, PSNR, and LPIPS performance of different schemes versus the number of active devices $K$ at SNR$=25$\,dB. Here, we plot ``Orth-Com" at $\text{BER}=0$ (error-free channel), hence, its TER equals zero and its PSNR/LPIPS performance represents the ideal performance without any token errors. With the increase of $K$, the TER of ``Non-Orth Com" increases due to the increased probability of token collisions. Hence, the PSNR and LPIPS performance of ``Non-Orth Com" degrades significantly as $K$ increase. Thanks to the masked token prediction from the candidate token sets, the proposed ToDMA experiences a much slower TER increase and its PSNR/LPIPS performance only slightly degrades. 

The visual quality results for \( K=40 \) are presented in Fig. \ref{Sim:Visual-1}, illustrating the impact of token collisions and the ability of ``ToDMA" to overcome them. To better highlight this comparison, ``ToDMA" and ``Non-Orth Com" are assumed to have perfect estimation of the token set \( \mathcal{P}_n \), while ``Orth-Com" operates with $\text{BER}=0$. It can be seen that the visual quality achieved by ``ToDMA" is quite similar to that achieved by ``Orth-Com", while there are obvious visible distortions in the images recovered by ``Non-Orth Com". This can also be seen from the token error visualizations, where each image's 256 tokens are represented on a 16×16 grid. Green and gray blocks indicate correct  and incorrect tokens, respectively. It is obvious that the token errors due to collisions in the ``Non-Orth Com" scheme can be significantly reduced by the proposed ``ToDMA" scheme.

\begin{figure}[t]
\vspace{-4mm}
     \centering
     \includegraphics[width = 0.65\columnwidth,keepaspectratio]{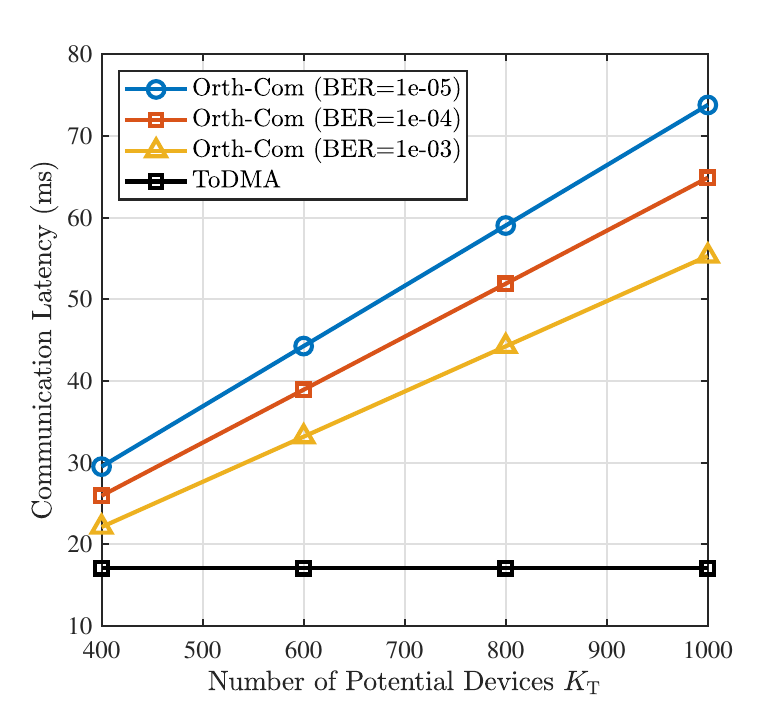}
     \captionsetup{font={footnotesize, color = {black}}, singlelinecheck = off, justification = justified,name={Fig.},labelsep=period}
     \caption{Latency compared to adaptive QAM with various target BERs.}
     \label{Sim:Latency}
     \vspace{-8mm}
\end{figure}

\subsection{Latency Comparisons}
To calculate the communication latency, we consider an OFDM system, where the number of subcarriers is set to $N_s=1024$ and the subcarrier spacing is $f_s = 15$\,kHz. For the Orth-Com benchmark, the entire bandwidth is uniformly assigned to $K_{\rm T}$ devices; hence, the rate of each device can be approximated as
\cite{goldsmith1997variable}
$R_{\text{Orth}} = \frac{N_s f_s}{K_{\rm T}}{\log _2}( 1 + \frac{1.5}{-\ln\left(5 \text{BER}\right)} \text{SNR}  )$.
Then, the communication latency of Orth-Com benchmark can be calculated as $N\log_2(Q)/R_{\text{Orth}}$. For the proposed ToDMA framework, the latency can be calculated as $LN/(N_sf_s)$, which has no relationship to $K_{\rm T}$. The numerical comparison at SNR=$25$\,dB is shown in Fig. \ref{Sim:Latency}. It can be seen that the latency of the proposed ToDMA is several times less than that of the Orth-Com scheme, especially for a larger total number of devices $K_{\rm T}$. 

\vspace{-2mm}
\section{Conclusions}
We introduced the novel ToDMA framework designed for massive access semantic communications in next-generation 6G networks. In ToDMA, a large number of devices utilize a shared token codebook derived from pre-trained foundation models to tokenize source signals into token sequences. Each token is then mapped to a codeword from a shared orthonormal modulation codebook. To support massive connectivity with low latency, ToDMA facilitates non-orthogonal grant-free transmission. At the receiver, active tokens and their associated CSI are estimated through correlation. Active tokens are then assigned to their corresponding devices based on CSI similarity. For the unassigned tokens, i.e., collided tokens, we leverage a pre-trained bi-directional transformer to associate them to the masked positions in the reconstructed token sequences, using the context and semantic orthogonality. ToDMA can be considered as a joint source-channel coding scheme, where the correlations among source token sequences are leveraged to recover collisions over the channel. Simulation results demonstrate that ToDMA outperforms context-unaware orthogonal and non-orthogonal communication schemes in both communication latency and image transmission quality.



\end{document}